\documentclass[a4paper]{article}

\usepackage[utf8]{inputenc}
\usepackage[british]{babel}
\usepackage{hyperref}
\usepackage{csquotes}
\usepackage[sorting=none,bibstyle=numeric,citestyle=numeric-comp,maxbibnames=20]{biblatex}

\usepackage{fourier}
\usepackage{microtype}

\usepackage{algorithm}
\usepackage{algpseudocode}

\usepackage{booktabs}
\usepackage{tabularx}

\usepackage{amsmath}
\usepackage{amssymb}
\usepackage{siunitx}

\usepackage{svg}
\usepackage{subcaption}

\usepackage{authblk}

\addto\extrasbritish{%
}

\begin{document}

\title{Error estimation and step size control with minimal subsystem interfaces}
\author{Lars T. Kyllingstad}
\affil{SINTEF Ocean, Trondheim, Norway}
\author{Severin Sadjina}
\author{Stian Skjong}
\affil{SINTEF Nordvest, Ålesund, Norway}
\date{}
\maketitle

\begin{abstract}
    We review error estimation methods for co-simulation, in particular methods that are applicable when the subsystems provide minimal interfaces.
    By this, we mean that subsystems do not support rollback of time steps, do not output derivatives, and do not provide any other information about their internals besides the output variables that are required for coupling with other subsystems.
    Such ``black-box'' subsystems are common in industrial applications, and the ability to couple them and run large-system simulations is one of the major attractions of the co-simulation paradigm.
    We also describe how the resulting error indicators may be used to automatically control macro time step sizes to strike a good balance between simulation speed and accuracy.
    The various elements of the step size control algorithm are presented in pseudocode so that readers may implement them and test them in their own applications.
    We provide practicable advice on how to use error indicators to judge the quality of a co-simulation, how to avoid common pitfalls, and how to configure the step size control algorithm.
\end{abstract}


\section{Introduction} \label{sec:introduction}

\emph{Co-simulation} is a versatile method for simulating coupled dynamical systems.
It is most readily described by comparing it with the \textit{de facto} standard simulation method, which we will refer to as \emph{monolithic} simulation.\footnote{
    For a more precise and complete definition of co-simulation and its associated terminology, we refer to the review by \textcite{gomes2018cosimulation}.
    Note that we focus entirely on \emph{continuous-time} co-simulation in the present paper.
}

In a monolithic simulation, all the model equations are represented in one state space system and solved jointly by a single program: the \emph{solver}.
Often, the modelling language is proprietary and closely tied to a specific solver in a unified modelling and simulation environment.

Co-simulation, by contrast, does \emph{not} require all subsystems to share state space, or to use the same solver.
Each subsystem model is solved independently of the others, using whichever solver and language is most appropriate for its particular dynamics.
To remain coupled, the subsystems exchange information and synchronise their logical clocks regularly, at so-called \emph{synchronisation points}.
The logical time that passes between two synchronisation points is called a \emph{macro time step}; its inverse is the \emph{synchronisation frequency}.
Each subsystem solver is free to use internal (\emph{micro}) time steps which are shorter than the macro steps.

Regardless of simulation technique, there is always a trade-off between stability and accuracy on one side and simulation speed on the other.
The former are usually quantified by an \emph{error estimate}, while the latter is measured by the \emph{real-time factor} (RTF), which is the logical (simulation clock) time divided by the actual (wall clock) time.
In practical terms, optimising the performance of a simulation usually boils down to one of the following:
\begin{itemize}
    \item minimising errors while keeping the RTF within certain bounds
    \item maximising the RTF while keeping errors within certain bounds
\end{itemize}
The first is typical for real-time simulations, while the latter is usually what we want in analytical simulations.
In both cases, we need some way to estimate errors and some way to tune the simulation speed.
As we shall see, co-simulation poses particular challenges in both regards.

Our goal with this article is to provide readers with the knowledge and tools to optimise the performance of their simulations while leveraging one of co-simulation's greatest strengths:
the ability to couple subsystems while sharing a minimal amount of information between them---that is, to use \emph{minimal subsystem interfaces}.
Let us start by discussing \emph{why} we consider this to be a strength (and when it is not).

\subsection{Pros and cons of co-simulation} \label{sec:proscons}

The main difference between co-simulation and monolithic simulation lies in the amount of information each subsystem must share, and the nature of that information.
In a monolithic simulation, the solver requires access to the model equations of all subsystems to solve them jointly.
More precisely, it needs the values and derivatives of all subsystem states.
In a co-simulation, it is only necessary to expose the variables that are involved in interactions between subsystems, that is, the \emph{coupling variables}.
A typical example would be the force one subsystem exerts on another.

A slightly different take on this is that co-simulated systems are \emph{sampled systems}, in that the different subsystems sample each other's coupling variables.
Monolithic simulations more closely approximate continuous systems.

In software-architectural terms, the modules that represent subsystems in a co-simulation are \emph{loosely coupled}, since they exchange relatively little information, and do so comparatively rarely, through a narrow interface.
This modularity has many benefits.
Subsystem models can more easily be reused in different contexts, they can more readily be combined into larger coupled systems, and they can be shared with less fear of exposing sensitive information.
Solver characteristics, such as integration method and the length of micro time steps, can be tailored to the particular dynamics of each subsystem.
Co-simulation lends itself well to distributed simulation, meaning that the workload can be shared by multiple CPU cores or computers, and information security can be further enhanced through isolation of the running code.
This makes the method well suited for large and complex systems, for hardware-in-the-loop simulations, and for collaborative simulation-building---even between competing companies.

All of the above have contributed to the method gaining popularity in several industries, notably the automotive, energy, and robotics industries, and more recently in the maritime sector \cite[][see references therein]{gomes2018cosimulation}.
Our own research has mainly focused on the latter \cite{sadjina2019distributed, skjong2018virtual}.

However, the loosely coupled nature of co-simulation can also be a major drawback, as it makes the method less suited for systems that are tightly coupled in the \emph{dynamical} sense.
In principle, by the Nyquist--Shannon sampling theorem, the synchronisation rate of two subsystems need only exceed the Nyquist rate of the signals exchanged between them to avoid aliasing.
In practice, the synchronisation rate must usually be much higher, due to the use of low-order signal reconstruction techniques inside subsystems.
A higher synchronisation rate means more frequent communication, and thus reduced simulation speed.\footnote{
    One might ask why a co-simulation should take a bigger performance hit from a small step size than a monolithic simulation would.
    The answer is that monolithic simulation software has the benefit of being able to keep all variables in a shared memory space accessible by the solver.
    Co-simulation software is by nature designed to support communication \emph{across} memory barriers---between different software modules, processes, platforms, or network entities---which is generally much slower than memory access.
    As the macro time step size decreases, we eventually enter an I/O-bound regime where the simulation time increases in inverse proportion.
}

Aliasing and low-order signal reconstruction, together with discontinuities at synchronisation points, are the main causes of what we call \emph{coupling errors} \cite{benedikt2019relaxing}.
These are errors which occur as a consequence of the coarse-grained discretisation of communication between subsystems in a co-simulation; they would not occur in a monolithic simulation.

\subsection{Dealing with coupling errors}
\label{sec:introduction_stepsize}

There are many strategies we can employ to deal with rapidly varying signals between subsystems and the coupling errors that arise as a consequence.
One is to keep tight couplings enclosed within subsystems.
In principle, this is a great idea.
Co-simulation and modularity are means to an end, not ends unto themselves; it makes no sense to split a system model into its smallest possible components just for the sake of it.
Unfortunately, there is no general rule as to what is the ``right'' way to do the splitting either.
Many factors come into play, often beyond the control of the person who runs the simulation at the end of the day.
They could be technical limitations, such as different modelling software being required for different subsystems; organisational limitations, for example that two subsystem models are supplied by two different companies; or financial reasons, for example if splitting subsystems in a certain way makes them more amenable for reuse, thereby saving labour costs in the long run.
The goal of avoiding tight couplings, while laudable, may very well find itself near the bottom of the list.

Other strategies to reduce errors include the use of higher-order signal reconstruction techniques inside subsystems, and \emph{restepping}: rollback of one or more macro time steps, usually to repeat them with adjusted inputs as part of an iterative algorithm or to split a long time step into several shorter ones.
Various forms of these approaches have been shown to work well in the research literature, both separately and in combination \cite{busch2019performance},
but they are often not applicable in practice.
While the major commercial modelling tools increasingly provide support for
such techniques, many applications still require the use of niche software and
handwritten models that don't.
This is especially true for multi-physical and cyber-physical systems, and it is
compounded by the fact that several of the subsystems in a co-simulation must
support the methods for them to be effective.

An interesting class of strategies is based on the idea of \emph{input corrections}.
Here, the co-simulation master algorithm intervenes and modifies signals in-flight between subsystems to compensate for coupling errors.
Examples based on energy conservation \cite{benedikt2013nepce, sadjina2016energy, benedikt2019relaxing, rodriguez2022energybased} and H$_\infty$ synthesis \cite{chen2022explicit} have demonstrated that it is possible to apply such corrections without requiring that subsystems support rollback or derivatives.
A deeper discussion of input correction methods is outside the scope of this paper, but we do want to note that they can be a powerful complement \cite{sadjina2020energy} to \emph{macro step size control}, which will be one of our main topics.

A radically different approach to error control is offered by \emph{transmission
line modelling} \cite{braun2022numerically}.
The idea here is that, since all physical signals have a finite propagation
speed, finitely separated real-world systems are literally independent of each
other at any given time.
This can be exploited when decoupling the systems for a co-simulation.
Capacitive elements at subsystem boundaries, such as springs or compressible
fluids, are replaced by \emph{transmission line elements} which are
modelled with wave propagation dynamics.
This decouples the subsystems in time, effectively replacing the numerical
coupling error with a modelling error which is arguably more manageable.
However, the macro step size is bounded from above by the signal propagation
delays \cite{braun2024transmission}, so care must be taken when choosing where to
place the subsystem boundaries.

Finally, we turn to \emph{step size control}, which, in a co-simulation context, refers to the continual adjustment of the length of macro time steps throughout a simulation.
The goal is generally to keep the system stable and within certain error bounds while maintaining a satisfactory simulation speed.

As we have already argued, the maximal step size is determined by the interaction dynamics of the subsystems:
Faster-varying signals require shorter time steps.
The majority of systems in real-world engineering are nonlinear, which means that their dynamical characteristics can change over time.
Such characteristics include both the signal frequencies and the subsystems' sensitivity to inputs.
When a lot is going on in the simulated system, we may have to reduce the step size to maintain accuracy, thereby reducing the simulation speed.
But when things calm down, we can increase the step size again to save time.

The key to knowing when to adjust the step size up or down, and by how much, is to have a good error estimate.
For a thorough discussion of co-simulation error estimation and step size control, see the 2021 paper by \textcite{meyer2021cosimulation}.
As they show, error estimation is to some extent independent of the particular step size control mechanism.
Like them, we will therefore treat these two topics separately in this paper.

\subsection{Minimal subsystem interfaces}

Now that we have discussed co-simulation in general, its strengths and weaknesses, and methods to deal with the latter, it is time to narrow our focus a bit.
As we've mentioned, one of the greatest strengths of the technique is the ability to encapsulate a vast variety of subsystems behind a narrow, well-defined interface.
A key point here is: The simpler the interface, the easier the subsystem is to build, use, and reuse.

We can distinguish between two layers of a subsystem's interface:
\begin{itemize}
    \item the \emph{programming interface}, which is defined by the functions made available for the co-simulation software to call, their names, their signatures, and their expected behaviour
    \item the \emph{model interface}, which is defined by the information communicated through the programming interface for a particular subsystem
\end{itemize}
An example of the former would be a function to set the value of input variables, while the latter would be the names and types of the variables (for example \emph{velocity}).
Over the last decade, the \emph{Functional Mock-up Interface} (FMI) has emerged as a \textit{de facto} programming interface standard for co-simulation \cite{junghanns2021fmi,schweiger2019empirical}.
Some more or less domain-specific efforts have been made to standardise model interfaces too \cite{irt2020mic,osp2020is,prostep2022smartse,asam2023osi}.

The \emph{minimal} interface required to use a subsystem in a co-simulation is indeed very simple.
On the programming interface level, it only needs to provide the master algorithm with the means to
\begin{itemize}
    \item set the values of input variables
    \item perform a macro time step of a certain length
    \item get the values of output variables
\end{itemize}
On the model interface level, it only needs to expose coupling variables and parameters.
Regardless of whether the subsystem is a mathematical model, a control system, or a piece of hardware, the information and functionality needed to support such an interface will already be present, so making it available to the co-simulator is relatively easy.

Many error estimation methods and master algorithms require additional features beyond this, as we shall see.
The most common such features are listed in \autoref{tab:advanced_features}.
We shall refer to these, and anything else beyond the minimal set, as \emph{advanced} features.
To use a subsystem in a co-simulation with advanced requirements, substantial work must usually be done by the implementer to provide the additional information and functionality.
Sometimes it is practically impossible, because the implementer of the interface is far removed from the person who develops the subsystem innards or uses the subsystem in a co-simulation.
For example, this could be the case if one uses a commercial modelling software to develop models that are exported for co-simulation.
\begin{table}
    \centering
    \begin{tabularx}{\textwidth}{>{\raggedright}p{2cm} >{\raggedright}X >{\raggedright}p{3cm}}
        \toprule
         Advanced feature &
            Description &
            Programming interface requirements
            \tabularnewline \addlinespace
         \midrule
         Rollback &
            Ability to restore the precise subsystem state at a previous synchronisation point. &
            Save state.\\Restore saved state.
            \tabularnewline \addlinespace
        Information exposed &
            Information about model structure or implementation beyond coupling variables (for example type and order of integration method, input extrapolation order, stiffness, or values of internal state variables). &
            Additional metadata.\\Get/set non-coupling variables.
            \tabularnewline \addlinespace
        Time derivatives &
            Derivatives of output variables with respect to time. &
            Get time derivatives.
            \tabularnewline \addlinespace
        Directional derivatives &
            Derivatives of output variables with respect to input variables. &
            Get directional derivatives.
            \tabularnewline \addlinespace
        Model adaptations &
            Changes to the model structure or implementation to make them suitable for a specific master algorithm. &
            (none)
            \tabularnewline \addlinespace
        \bottomrule
    \end{tabularx}
    \caption{%
        Examples of ``advanced'' subsystem features.
        While the last one does not necessarily impose additional \emph{interface} requirements, we include it in this list because it is perhaps the most invasive one from a subsystem modeller's perspective.%
    }
    \label{tab:advanced_features}
\end{table}

Our goal with this article is to enable scientists and engineers to enjoy the greatest benefits of co-simulation---reusability, scalability, information hiding, and more---while ensuring that their simulations are as accurate, stable, and fast as possible.
It addresses some of the most pressing challenges associated with co-simulation today, as identified in a survey by \textcite[][see Table 5 therein]{schweiger2019empirical}.
In particular, ``difficulties due to insufficient communication between theorists and practitioners'' are mitigated by the fact that minimal interfaces are easier to describe, and the algorithms that use them are easier to explain.
Our discussion about error estimation techniques in \autoref{sec:errorestimation} directly addresses common ``difficulties in judging the validity of a co-simulation''.
The combination of those techniques, the step size controller we will describe in \autoref{sec:stepsizecontrol}, and the practical advice we give in \autoref{sec:discussion} will hopefully help with ``difficulties in choosing the right co-simulation orchestration algorithm'', ``difficulties in how to define the macro step size for a specific co-simulation'', ``numerical stability issues of co-simulation'', and ``issues because of too simplistic extrapolation functions''.


\section{Error estimation}
\label{sec:errorestimation}

As mentioned in \autoref{sec:introduction_stepsize}, the ability to control the macro time step and, thus, the performance of a co-simulation requires the ability to estimate its accuracy according to some measure.
This raises two questions:
\begin{enumerate}
    \item How should we define such an accuracy or, conversely, error measure for co-simulations?
    \item How do we estimate its magnitude reliably without access to a reference solution?
\end{enumerate}
The main challenge in answering the first question is that we typically have no knowledge of the system states or the governing equations, either because subsystems are based on sensors, data-driven models, or even lookup tables, or because such information is not exposed via the subsystem interface. 
Instead, we define \emph{coupling errors} as errors which emerge solely because a given system model is co-simulated, and on top of errors which would also be present had the same simulation been carried out strictly monolithically.
To this end, we define the \emph{local} coupling error accrued during one macro time step $t[i-1] \rightarrow t[i]$ with respect to some quantity $y$ as
\begin{equation}
    \label{equ:error}
    \Delta y[i]
    \equiv
    y[i] 
    -
    {y_0}[i] 
    ,
\end{equation}
where $y[i]$ and ${y_0}[i]$ are the co-simulated solution and the monolithic reference solution, respectively, both at synchronisation point $i$, and it is assumed that $y[i-1] = {y_0}[i-1]$.
What this error represents precisely will depend on the specific system and co-simulation setup.
For example, it could represent a spurious position, force, or energy.

Here, we should distinguish between \emph{physical} couplings, which represent physical interactions between subsystems, and what we might call \emph{digital} connections, which represent information channels.
Examples of the latter include control signals, signals from sensors, and interactions with human interface devices.
Only physical couplings should be subject to error estimation of the kind we are discussing here.
Digital connections carry what would be sampled signals in the real world too, and the subsystems that send or receive them are designed to deal with them as such.
Therefore, we are not making the same kind of discretisation error when co-simulating them as we do for physical couplings.
(That's not to say that we are not making errors, but avoiding them is more a matter of ensuring that the synchronisation frequency is compatible with the sampling rate ranges expected by the subsystems in
question.)

The second question about the quantification of errors in co-simulation without knowledge of a reference solution $y_0$ is what the remainder of the present section is devoted to.

\subsection{A simple example}
\label{subsec:errorestimation_example}

Before venturing on, let us first consider the (very) simple but illustrative example of a mass--spring harmonic oscillator.
With no dissipation present, this is a closed system and its energy will be conserved over time.
Any arbitrary initial condition for its position and velocity results in a periodic sinusoidal motion which would persist \emph{ad infinitum}, leaving the total energy of the system unaltered.
Any suitable and correctly configured monolithic solver worth its salt should be able to simulate such a system with reasonable accuracy.
But if we, instead, consider the case of a non-iterative co-simulation coupling with mass and spring split into separate models with their own solvers each, we can observe a general discrepancy with respect to the system state.
Any \emph{finite} macro time step will introduce a lag in the communication of the coupling signals.
In this simple example, such delays will, over time, result in a diverging system state and an unstable co-simulation.
Effectively, the conservation of the total system energy is violated because spurious energy is added to it at each macro time step.
Such errors are solely a consequence of the co-simulation coupling and are absent in a monolithic simulation of the same system.

To illustrate this point let us consider a mass (subsystem $S_1$) with $m = \SI{100}{\kilogram}$ and a spring (subsystem $S_2$) with $k = \SI{1e3}{\newton\per\meter}$.
Both subsystems are stepped in parallel (Jacobi coupling scheme) and synchronised with a (deliberately large) macro time step of $\Delta t = \SI{0.05}{\second}$.
The time integration inside the spring subsystem is performed with the forward Euler method and a micro time step of $\delta t_2 = \SI{1e-4}{\second}$.
The system is initialised with velocity $v(t=0) = \SI{0}{\meter\per\second}$ and position $x(t=0) = \SI{1}{\meter}$ for the mass, that is, in a state where an energy of $E(t=0) = \SI{500}{\joule}$ is stored in the spring.
Because we would like to consider energies for our example, we model the physical connection between both subsystems as a \emph{power bond} so that the product of input and output signals represents a physical power:
The spring outputs the force $y_2 = F$ determined by the position of the mass, and the mass outputs its velocity $y_1 = v$ subject to the force exerted upon it.

\begin{figure}
    \centering
    \begin{subfigure}{\textwidth}
        \centering
        \includesvg[width=\textwidth]{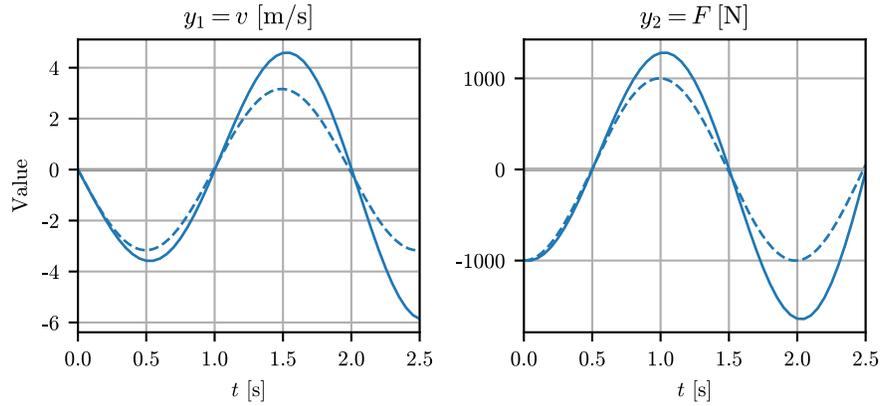}
        \caption{Subsystem outputs: mass velocity $y_1 = v$ and spring force $y_2 = F$.}
        \label{fig:toy_system_errors_outputs}
    \end{subfigure}
    \begin{subfigure}{\textwidth}
        \centering
        \includesvg[width=\textwidth]{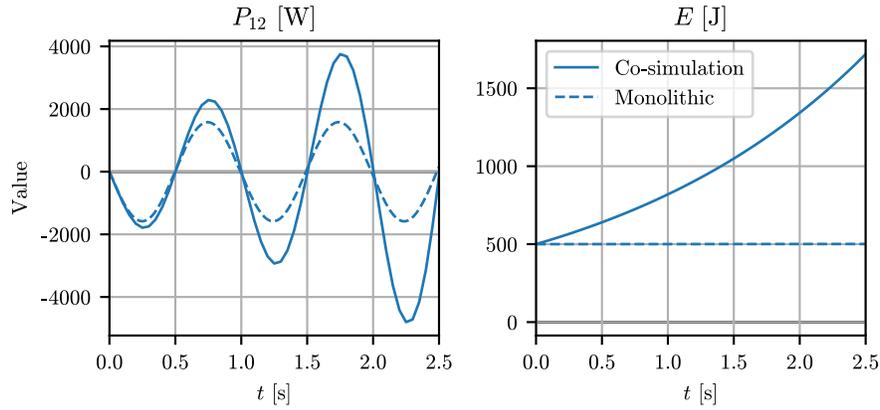}
        \caption{Power $P_{12}$ transmitted from mass $S_1$ to spring $S_2$ and total system energy $E$.}
        \label{fig:toy_system_errors_powers}
    \end{subfigure}
    \caption{%
    (\subref{fig:toy_system_errors_outputs}) Subsystem outputs and~(\subref{fig:toy_system_errors_powers}) transmitted power and total system energy for a 1-D mass--spring harmonic oscillator.
    There is a lag in the co-simulated results (solid) compared to the monolithic simulation (dashed), and this has important consequences:
    The output values are consistently too large in magnitude, and so is the flow of energy between the subsystems.
    This erroneously inflates the system's energy, distorting its dynamics and destabilising it as a manifestation of co-simulation errors.%
    }
    \label{fig:toy_system_errors}
\end{figure}

Now we are all set to compare the co-simulation of this harmonic oscillator to the monolithic reference simulation, which is also performed with forward Euler and a step size of $\delta t = \SI{1e-4}{\second}$.
As can be observed from \autoref{fig:toy_system_errors}, co-simulation introduces a constant time lag which produces velocity and force values that are constantly too large in magnitude.
Now it also becomes clearer why we chose to use a power bond to model the interaction between mass and spring:
It allows us to calculate the power transmitted between both subsystems from the product of their output signals, $|P_{12}| = |y_1 y_2| = |v| |F|$.
And with both velocity and force being too large in magnitude, co-simulation coupling continuously feeds spurious energy into the system.

\begin{figure}
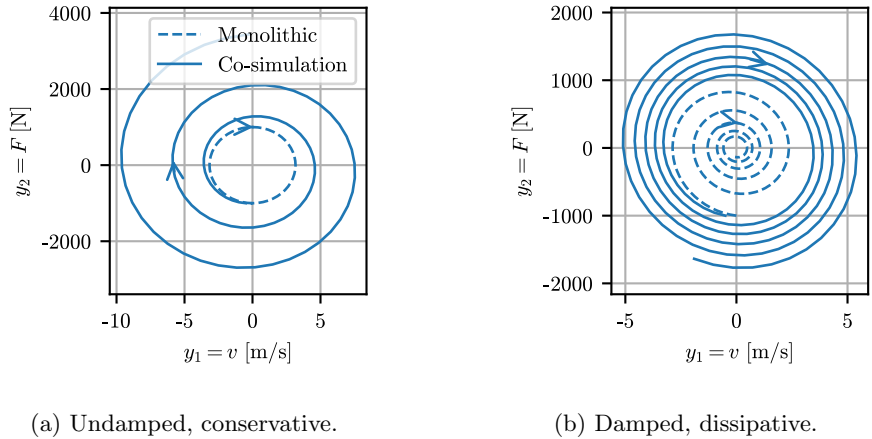

    \centering
    \begin{subfigure}{0.45\textwidth}
        \centering
        \includesvg[width=1\textwidth]{figures/phase_diagram_undamped.svg}
        \caption{Undamped, conservative.}
        \label{fig:toy_system_phase_diagrams_undamped}
    \end{subfigure}
    \hfill
    \begin{subfigure}{0.45\textwidth}
        \centering
        \includesvg[width=1\textwidth]{figures/phase_diagram_damped.svg}
        \caption{Damped, dissipative.}
        \label{fig:toy_system_phase_diagrams_damped}
    \end{subfigure}
    \caption{%
        Phase space diagram of subsystem outputs for a 1-D mass--spring harmonic oscillator.
        Without damping~(\subref{fig:toy_system_phase_diagrams_undamped}), the system is unstable for any finite macro step sizes (marginally stable), and the co-simulation (solid) diverges from the stable monolithic simulation (dashed).
        But even with damping present~(\subref{fig:toy_system_phase_diagrams_damped}), the co-simulation can still become unstable.%
    }
    \label{fig:toy_system_phase_diagrams}
\end{figure}

Before moving on to discussing various co-simulation error estimation methods, allow us to make one last point:
The undamped harmonic oscillator is, in fact, ill-suited for co-simulation because it is marginally stable, that is, only in the limit of $\Delta t \rightarrow 0$.\footnote{
    The observant reader may have noticed that the exact same is true also for the internal forward Euler time integration carried out in the spring subsystem, which is only stable in the limit $\delta t \rightarrow 0$.
    We chose it regardless because of its simplicity and because its instability has no consequence for the short time period we consider here.
}
We will continue using it for illustrative purposes, but we should replace it with a \emph{damped} harmonic oscillator to make it stable in principle.
We do so by including a linear damping element with a damping constant of $d = \SI{40}{\newton\second\per\meter}$ with the spring subsystem $S_2$.
Note that a co-simulation of the harmonic oscillator can still be unstable even with dissipation present, as illustrated in \autoref{fig:toy_system_phase_diagrams}.

\subsection{Error estimation methods}
\label{subsec:errorestimation_estimation}

Several proposals for co-simulation error estimation methods can be found in the literature.
Most of these are direct adaptations of methods from ordinary differential equation (ODE) and differential-algebraic equation (DAE) time integration schemes.
There, errors can be defined directly in terms of the system states, which is probably the most straightforward and natural way.
Co-simulation typically does not afford us the luxury of exposed system states, however, and most co-simulation error estimation methods consider errors in the model outputs (or derivations thereof) instead.
When it comes to the adaptation of schemes from classic numerical integration, three approaches have been taken\ \cite{gomes2018cosimulation, meyer2021cosimulation}:
\begin{description}
    \item[Embedded (or local extrapolation) methods] compare two co-simulation results of different (polynomial) order.
    \item[Milne's device] compares two co-simulation results of the same (polynomial) order, but using different integration methods.
    \item[Richardson extrapolation] compares two co-simulation results obtained with different macro time step sizes.
\end{description}
While their application to co-simulation is straightforward and comes with a substantial amount of historical background, they all fall short when it comes to requiring rollback, or at least the parallel execution of a second co-simulation.
By contrast, there exist methods which pose few requirements to the subsystems and their interfaces.
\autoref{tab:error_estimation_methods_overview} provides an overview of co-simulation error estimation methods and their requirements.
\begin{table}
    \centering
    \scriptsize
    \begin{tabularx}{\textwidth}{>{\raggedright}X >{\raggedright}p{0.25cm} >{\raggedright}p{0.25cm} >{\raggedright}p{0.25cm} >{\raggedright}p{0.25cm} >{\raggedright}p{0.45cm} >{\raggedright}p{3.25cm}}
        \toprule
        Error estimation method & \multicolumn{4}{c}{Advanced feature} &  & Notes \tabularnewline
        \cmidrule{2-5}
        & \rotatebox{90}{Rollback} & \rotatebox{90}{Information exposed} & \rotatebox{90}{Derivatives} & \rotatebox{90}{Model adaptations} & \rotatebox{90}{Number of parameters} & \tabularnewline
        \midrule
        Local input error estimation \cite{benedikt2013nepce,sadjina2020energy}: Comparing inputs against extrapolated inputs &  & $\circ$ &  $\circ$ &  & 0 & May require directional derivatives if direct feed-trough is present. \tabularnewline
        \addlinespace
        Explicit predictor/corrector \cite{busch2012effizienten, eguillon_f3ornits_2022}: Comparing outputs against predictions &  & $\times$ & & $\circ$ & $N_y$ & Requires input extrapolation orders. Issues for large step sizes. \tabularnewline
        \addlinespace
        Energy residuals (ECCO) \cite{sadjina2016energy, sadjina2024energy}: Comparing energy flows between models using power bonds &  & &  & $\circ$ & 0 & Estimates error in energy balance. Can also be used without power bonds. \tabularnewline
        \midrule
        Embedded method \cite{gunther2001multirate}: Comparing input extrapolations with different polynomial orders & $\times$ & $\times$ & $\circ$ & $\circ$ & $N_y$ & Requires control over polynomial input extrapolation. \tabularnewline
        \addlinespace
        Milne's device \cite{milne1926numerical, busch2012effizienten, meyer2021cosimulation}: Comparing integration with different methods of same order & $\times$ & $\times$ &  & $\circ$ & $N_\text{S}$ & Requires control over integration schemes. \tabularnewline
        \addlinespace
        Richardson extrapolation \cite{schierz2012cosimulation, arnold2013error}: Comparing results using different macro step sizes & $\times$ &  & &  & 0 & Computationally expensive, but can be optimised. \tabularnewline
        \addlinespace
        Compound-Fast method \cite{verhoeven2008bdf}: Step stiffer of two subsystems again with smaller step size and compare & $\times$ & $\times$ &  &  & $\geq 1$ & Requires knowledge about subsystem dynamics and proper choice of step sizes. \tabularnewline
        \addlinespace
        Error Differential Equation \cite{genser2022approximated} to approximate extrapolation error &  &  & $\times$ &  & $\geq 0$ & Requires (approximated) directional derivatives. \tabularnewline
        \bottomrule
    \end{tabularx}
    \caption{%
        Overview of co-simulation error estimation methods and the requirements they place on models, model interfaces, modellers, and users in terms of the advanced features listed in \autoref{tab:advanced_features}.
        A cross ($\times$) denotes that a feature is required to use the method and a circle ($\circ$) denotes that the method may sometimes rely on or benefit from the feature.
        $N_\text{S}$ denotes the number of subsystems and $N_y$ the number of subsystem outputs to which the method is applied.
        The three topmost methods, which we consider the ones that pose the fewest requirements, are described in greater detail in \autoref{subsec:errorestimation_estimation_minimal}.%
    }
    \label{tab:error_estimation_methods_overview}
\end{table}

\subsection{Error estimation with minimal requirements}
\label{subsec:errorestimation_estimation_minimal}

Here, our focus is on approaches which are relatively easy to understand, use, and implement, and which place minimal requirements on models and model interfaces.
We specifically regard this to be the case for methods which do not require repetition of macro time steps (\emph{rollback}) or derivatives (of interface signals with respect to time or input signals), and which come with a minimum number of parameters to set and tune.
Consequently, we will focus on the three topmost entries in \autoref{tab:error_estimation_methods_overview} in the present section.
Note that we do not wish to provide a \emph{performance} comparison between these methods, and the simple case discussed here should not be used as an indication for how accurately the methods estimate errors.
Instead, we aim to describe when and how they can be used, how to implement them, and what advantages and disadvantages they bring about.
For transparency, note that the third method, \emph{ECCO}, was developed by the authors.

\subsubsection{Local input error estimation (NEPCE)}
\label{subsec:errorestimation_estimation_nepce}

In co-simulation, inputs are only synchronised across coupled subsystems at discrete communication time points.
Consequently, they have to be extrapolated while the subsystems advance in time toward the next communication point.
Often input values are simply held constant, and this is called \emph{zero-order hold} (ZOH).
Either way, this lack of synchronization and extrapolation---whether stemming from a deliberate effort to approximate input values, or not---leads to local errors in the subsystem inputs.

The \emph{nearly energy-preserving coupling element} (NEPCE)\cite{benedikt2013nepce} was proposed to correct for such errors by altering input values before they are set for the next time step.
Input corrections---while very much worth looking into in their own right---are beyond the scope of the present work.
Nonetheless, NEPCE gives us a straightforward and simple means to estimate coupling errors that only uses the input signals:
\begin{equation}
\label{equ:error_nepce}
    \Delta \textbf{u}(t)
    \approx
    \textbf{u}(t) - \tilde{\textbf{u}}(t)
    .
\end{equation}
If any of the outputs depend directly on any of the inputs---in other words, if \emph{direct feed-through} is present---an additional term should be included~\cite{sadjina2020energy},
\begin{equation}
\label{equ:error_nepce_modified}
    \Delta \textbf{u}(t)
    \approx
    \big(
    1 - \mathrm{L} \mathrm{J}_y(\mathbf{u})
    \big)^{-1}
    \big(
    \textbf{u}(t) - \tilde{\textbf{u}}(t)
    \big)
    ,
\end{equation}
where $\mathrm{L}$ is the connection graph matrix which relates outputs to inputs, $\textbf{u} = \mathrm{L} \textbf{y}$, and ${J_y}_{ij} \equiv \partial y_i / \partial u_j$ is the interface Jacobian (directional derivatives).
The latter may not be available, but one is free to approximate it or to disregard it using \autoref{equ:error_nepce} without modification at the cost of estimation accuracy.
Alternatively, it may be possible to split the system in a way which does not lead to direct feed-through.
\begin{figure}
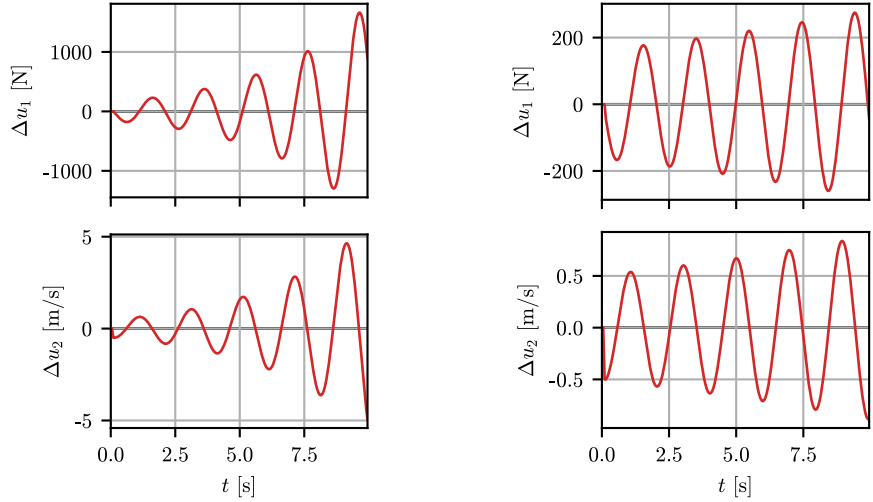

    \centering
    \begin{subfigure}{0.45\textwidth}
        \centering
        \includesvg[width=1\textwidth]{figures/error_estimate_nepce_undamped.svg}
        \caption{Undamped, conservative.}
        \label{fig:error_estimate_nepce_undamped}
    \end{subfigure}
    \hfill
    \begin{subfigure}{0.45\textwidth}
        \centering
        \includesvg[width=1\textwidth]{figures/error_estimate_nepce_damped.svg}
        \caption{Damped, dissipative.}
        \label{fig:error_estimate_nepce_damped}
    \end{subfigure}
    \caption{%
        Local input error estimation according to the \emph{nearly energy-preserving coupling element} (NEPCE) \cite{benedikt2013nepce} for the harmonic oscillator co-simulation example from \autoref{subsec:errorestimation_example} without~(\subref{fig:error_estimate_nepce_undamped}) and with~(\subref{fig:error_estimate_nepce_damped}) damping.%
    }
    \label{fig:error_estimate_nepce}
\end{figure}

\autoref{fig:error_estimate_nepce} demonstrates the application of local input error estimation for ZOH for which~\autoref{equ:error_nepce} simply becomes
\begin{equation}
\label{equ:error_nepce_zoh}
    \Delta \textbf{u}[i]
    \approx
    \textbf{u}[i] - \textbf{u}[i-1]
    .
\end{equation}
The spring--damper subsystem $S_2$ exhibits direct feed-through because its output depends directly on its input, ${J_y}_{22} = \partial y_2 / \partial u_2 = d$.
The difference is negligible for the examples considered here, but becomes significant for strong damping.

\subsubsection{Explicit predictor/corrector}
\label{subsec:errorestimation_estimation_predictor_corrector}

\begin{figure}
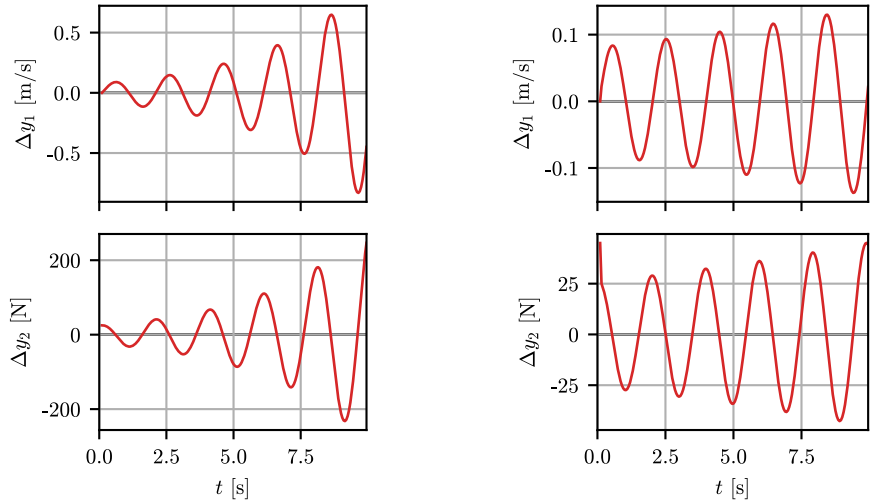

    \centering
    \begin{subfigure}{0.45\textwidth}
        \centering
        \includesvg[width=1\textwidth]{figures/error_estimate_predictor_undamped.svg}
        \caption{Undamped, conservative.}
        \label{fig:error_estimate_predictor_undamped}
    \end{subfigure}
    \hfill
    \begin{subfigure}{0.45\textwidth}
        \centering
        \includesvg[width=1\textwidth]{figures/error_estimate_predictor_damped.svg}
        \caption{Damped, dissipative.}
        \label{fig:error_estimate_predictor_damped}
    \end{subfigure}
    \caption{%
        Local output error estimation based on an \emph{explicit predictor/corrector} approach~\cite{busch2012effizienten} for the harmonic oscillator co-simulation example from \autoref{subsec:errorestimation_example} without~(\subref{fig:error_estimate_nepce_undamped}) and with~(\subref{fig:error_estimate_nepce_damped}) damping.%
    }
    \label{fig:error_estimate_predictor}
\end{figure}

If we know the order $m$ of the input extrapolation method used inside of a subsystem, we can compare the corresponding output value $y_k$ against a Lagrange polynomial prediction of order $r = m + 1$,
\begin{subequations}
\label{equ:error_predictor_prediction}
\begin{equation}
    \tilde{y}_k^{(r)}[i+1]
    =
    \sum_{j = 0}^{r} y_k[i-r+j] L_j^{(r)}\big(t[i+1]\big)
    ,
\end{equation}
where
\begin{equation}
    L_j^{(r)}(t)
    =
    \prod_{l=0, l\neq j}^r \frac{ t - t[i-r+l] }{ t[i-r+j] - t[i-r+l] }
    ,
\end{equation}
\end{subequations}
to obtain an estimate
\begin{equation}
\label{equ:error_predictor_error}
    \Delta y_k[i]
    \approx
    y_k[i]
    -
    \tilde{y}_k^{(r)}[i]
\end{equation}
for the local output error~\cite{busch2012effizienten}.
\autoref{fig:error_estimate_predictor} illustrates this estimation method for ZOH ($m = 0$) and fixed macro step sizes for which \autoref{equ:error_predictor_error} yields
\begin{equation}
\label{equ:error_predictor_error_zoh_fixed}
    \Delta y_k[i]  
    \approx
    y_k[i] - 2 y_k[i-1] + y_k[i-2]
    .
\end{equation}
Alternatively, constrained least squares can be used to obtain a prediction to compare against~\cite{eguillon_f3ornits_2022}.

\subsubsection{Energy-conservation-based co-simulation method (ECCO)}
\label{subsec:errorestimation_estimation_ecco}

Co-simulations are typically concerned with highly heterogeneous and modular systems in which energy storage, creation, and dissipation are modelled within subsystems, while energy fluxes between them are realised through their couplings.
We will naturally have a hard time trying to enforce energy conservation \emph{inside} of a subsystem without its extensive modification,
but we can aim for accurate energy fluxes \emph{between} subsystems.
This is the basic idea behind the \emph{energy-conservation-based co-simulation method} (ECCO) \cite{sadjina2016energy}, which uses \emph{power bonds} to model the energetic interactions between subsystems.
Power bonds, in this context, are input--output pairs whose product gives a physical power.
Examples include electric voltage--current couplings or the mechanical force--velocity coupling we have already encountered in \autoref{subsec:errorestimation_example}.

\begin{figure}
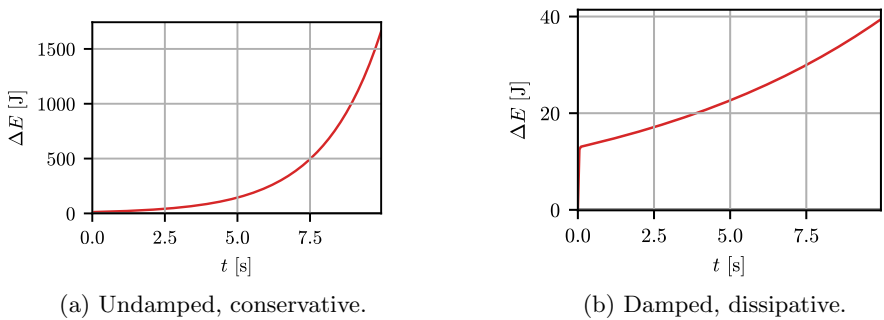

    \centering
    \begin{subfigure}{0.45\textwidth}
        \centering
        \includesvg[width=1\textwidth]{figures/error_estimate_ecco_undamped.svg}
        \caption{Undamped, conservative.}
        \label{fig:error_estimate_ecco_undamped}
    \end{subfigure}
    \hfill
    \begin{subfigure}{0.45\textwidth}
        \centering
        \includesvg[width=1\textwidth]{figures/error_estimate_ecco_damped.svg}
        \caption{Damped, dissipative.}
        \label{fig:error_estimate_ecco_damped}
    \end{subfigure}
    \caption{%
        Local energy error estimation according to the \emph{energy-conservation-based co-simulation method}~\cite{sadjina2016energy} (ECCO) for the harmonic oscillator co-simulation example from \autoref{subsec:errorestimation_example} without~(\subref{fig:error_estimate_nepce_undamped}) and with~(\subref{fig:error_estimate_nepce_damped}) damping.
    }
    \label{fig:error_estimate_ecco}
\end{figure}

For a given power bond, ECCO calculates a \emph{residual power} by summing over the individual powers $P_k$ transmitted by subsystem $S_k$ which are given as product of output $y_k$ and (extrapolated) input $\tilde{u}_k$,
\begin{equation}
\label{equ:error_ecco_residual_power}
    \delta P[i]
    \equiv
    -\sum_k P_k[i]
    =
    -\sum_k y_k[i] \tilde{u}_k[i]
    .
\end{equation}
The sign is chosen by definition, such that energy is erroneously added (leaked) when $\delta P > 0$ ($\delta P < 0$).
These residuals are spurious energy flows which stem from the fact that energy is \emph{not} conserved in the interactions between subsystems in a co-simulation, just as we have already discussed in \autoref{subsec:errorestimation_example}.
Simply put, because input values have to be extrapolated during macro time steps, and will, thus, deviate from the corresponding output values of a connected subsystem, the amount of energy that is transferred between them is not fully accounted for either:
One subsystem may transmit more energy than the other receives, or vice versa.
The method has two unique characteristics that are worth mentioning:
\begin{itemize}
    \item It considers errors per \emph{coupling}, and not per subsystem.
    \item The residual power is \emph{not} an estimate, but precisely the instantaneous rate of energy being erroneously added to or drained away from the system at the synchronisation point.
\end{itemize}
In terms of error estimation, we are interested in the amount of energy wrongfully added to, or taken from, the total system due to co-simulation coupling during one macro time step $t[i] \rightarrow t[i+1] = t[i] + \Delta t[i]$.
We obtain such a \emph{residual energy} by integrating \autoref{equ:error_ecco_residual_power},~\cite{sadjina2024energy}
\begin{equation}
\label{equ:error_ecco_energy_error}
    \delta E[i + 1]
    =
    \int_{t[i]}^{t[i+1]}
    \delta P(t)
    \mathrm{d}t
    \approx
    \frac{1}{m + 2}
    \delta P_k[i+1]
    \Delta t[i]
    .
\end{equation}
\autoref{fig:error_estimate_ecco} illustrates the relationship between the residual energy and the local error in the total system energy for ZOH ($m = 0$).

One advantage of this formalism is that the powers and energies are all additive.
This means, for example, that we can simply sum over the individual energy residuals for all couplings to obtain the total energy coupling error for an entire co-simulation.
Finally, note that this method also works on interfaces that are \emph{not} power bonds, but conservation arguments need not apply then.

\subsubsection{Comparison}

As we have seen, there exist several methods for estimating coupling errors in co-simulations.
While most require the use or implementation of one or several of the advanced features listed in \autoref{tab:advanced_features}, we picked and discussed three methods which we believe can be classified as placing minimal requirements on models, model interfaces, modellers, and users:
NECPE to estimate local input errors (\autoref{subsec:errorestimation_estimation_nepce}), the explicit predictor/corrector method to estimate local output errors (\autoref{subsec:errorestimation_estimation_predictor_corrector}), and ECCO to calculate local energy errors (\autoref{subsec:errorestimation_estimation_ecco}).
Each of these comes with their own unique set of characteristics that determine which one would be best suited for a given case, and it seems impossible to claim that any one is better than another in general.
What is common to them all is that none require rollback or incur a significant computational cost.

NEPCE is rather versatile and suitable for most cases because it requires minimal information about the subsystems, has no parameters to set or tune, and is simple to understand and to implement.
Its only downside is that it can become unreliable in the presence of direct feed-through, that is, when any of the outputs depend directly on any of the inputs.
This means that one should ideally check that no direct feed-through is present in any of the subsystems and, if there is, either implement the modified calculation given in \autoref{equ:error_nepce_modified} or alter the models to avoid direct feed-through altogether.
The explicit predictor/corrector method only requires output signals to calculate the predictor based on polynomials or by using constrained least squares.
It does, however, require knowledge of the input extrapolation orders used, although these may be inferred via the FMI API in principle~\cite{eguillon_f3ornits_2022}.
It also comes with somewhat more involved calculations and requires sufficiently small step sizes to work well~\cite{busch2012effizienten}.
Finally, ECCO requires no information from the subsystems, has no parameters to set or tune, and is comparatively easy to implement and use.
Because it is designed for use with power bonds it may require adaptations to existing subsystems, although it can be applied to any type of coupling.
If individual errors per interface signal are desired (instead of per coupling) or when power bonds are difficult to implement, other methods may be better suited.

\subsection{Construction of an error indicator}
\label{sec:error_indicator}

Now that we have the ability to estimate local errors, we will often be interested in aggregating them into a single \emph{error indicator}.
This is useful in its own right to gauge the performance of a co-simulation more easily, but it will also prove useful for automatic control of the macro step size.
There are infinitely many ways to realise such a mapping from a set of individual errors $\{\Delta y_1, \dots, \Delta y_{N_y}\}$ to one summarising scalar metric $\epsilon_y$, so let us here attempt to present a few common ones along with supplementary advice and remarks.

Typically, the individual errors will have to be normalised first.
This is usually done using absolute and relative tolerances, $\delta$ and $\sigma$, respectively,
\begin{equation}
\label{equ:error_indicator_normalize}
    \epsilon_{y_k}[i]
    \equiv
    \frac{
        \Delta y_k[i]
    }{
        \delta_k
        +
        \sigma_k
        \big| y_k[i] \big|
    }
    ,
\end{equation}
where $\Delta y_k$ is the estimated local error in $y_k$.
Tolerances should ideally be set for each quantity individually to reflect its desired resolution.
Absolute tolerances help avoid numerical issues related to small values of $|y|$.
Using position errors as an example, they could be set in relation to typical position scales of the problem at hand.
For our simple toy system from \autoref{subsec:errorestimation_example} an absolute position tolerance could be set to a fraction of the initial displacement of the mass.
Appropriately choosing tolerance values can be challenging because it requires a sufficient understanding of the system, its dynamics, and the interactions between the various subsystems.
Consequently, tolerances will often have to be guessed or determined via trial and error.

A few common choices for constructing scalar error indicators from individual (normalised) errors are the root-mean-squared error (RMSE),
\begin{subequations}
\label{equ:error_indicators}
\begin{equation}
\label{equ:error_indicators_rmse}
    \epsilon_\text{RMSE}[i]
    \equiv
    \sqrt{
        \frac{1}{N_y}
        \sum_{k = 1}^{N_y}
        \epsilon_{y_k}[i]^2
    }
    ,
\end{equation}
the mean absolute error (MAE),
\begin{equation}
\label{equ:error_indicators_mae}
    \epsilon_\text{MAE}[i]
    \equiv
    \frac{1}{N_y}
    \sum_{k = 1}^{N_y}
    \big|
        \epsilon_{y_k}[i]
    \big|
    ,
\end{equation}
or the use of a maximum function,
\begin{equation}
\label{equ:error_indicators_maximum}
    \epsilon_\text{MAX}[i]
    \equiv
    \max_k{
        \epsilon_{y_k}[i]
    }
    ,
\end{equation}
\end{subequations}
with many more available as direct applications from statistics or machine learning.
Each choice comes with their own strength and weaknesses (more on this in \autoref{subsec:error_recommendations_sensitivity}), but regardless of which function is chosen, its performance will depend on the co-simulation, the desired focus, and especially on the way the individual error contributions are normalised.

\subsection{Issues, pitfalls, and recommendations}
\label{sec:error_recommendations}

To round this section off, allow us to delve into some issues related to estimating errors in co-simulations.

\subsubsection{Reliability under large-error regimes}
\label{subsec:error_recommendations_large_errors}

\begin{figure}
    \centering
    \includesvg[width=\textwidth]{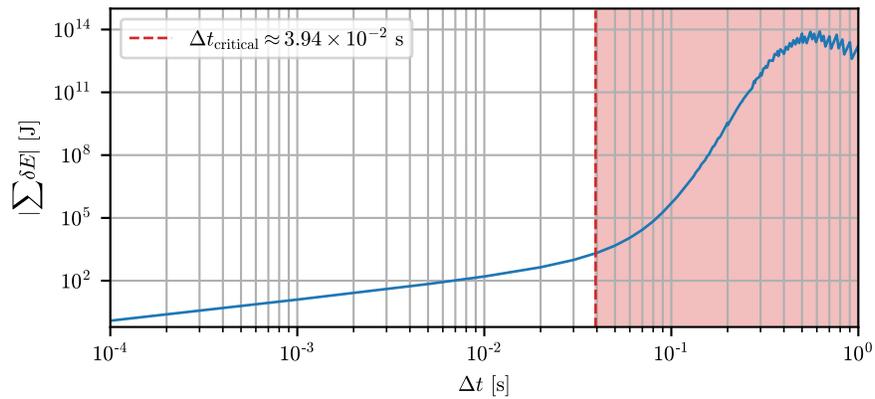}
    \caption{
    Relationship between macro step size and cumulative coupling error for the damped harmonic oscillator co-simulation example from \autoref{subsec:errorestimation_example}.
    Each data point is a separate co-simulation run with a fixed step size $\Delta t$.
    As suggested by \textcite{rahikainen2020automated}, such investigations can give strong indications for the stability of the co-simulation at different step sizes, as well as for the reliability of the error estimator used.
    Here, \autoref{equ:error_ecco_energy_error} is summed over the entire simulation time to estimate the cumulative error in the system's total energy.
    The critical step size (shown in red) beyond which the co-simulation becomes unstable lies on the transition between an approximately linear regime and one in which energy errors diverge quickly with increasing step sizes.
    On the other hand, the approximation of the energy error becomes unreliable for very large step sizes when oscillations set in.%
    }
    \label{fig:ecco_stability}
\end{figure}

One practical issue with many, if not most, of the suggested methods is that they tend to become less reliable with increasing errors.
This is rather unfortunate, of course, as the benefits of error estimation and control are greatest precisely under large-error regimes.
To gauge whether this is an issue, one should run repeated experiments with the same co-simulation using different macro step sizes.
Coupling errors should generally increase with increasing macro step size.
If, beyond a certain point, the reported error estimate \emph{decreases} with increasing macro step size, it is fair to assume that the step size will have to be kept below that level.
\autoref{fig:ecco_stability} illustrates this approach.

\subsubsection{Sensitivity to large numeric values}
\label{subsec:error_recommendations_sensitivity}

Another practical issue that seems to be mostly unnoticed in the literature and, thus, is present in most proposed error estimators and error indicators is the direct use of unnormalised subsystem outputs and errors~\cite{sadjina2016energy, sadjina2020energy}.
As explained in \autoref{sec:error_indicator}, different local errors are often compounded into a scalar error indicator across subsystems for the purpose of error analysis and automatic macro step size control.
Most methods suggest the use of the RMSE (\autoref{equ:error_indicators_rmse}) or a maximum function (\autoref{equ:error_indicators_maximum}), but both are very sensitive to large numeric values which can dominate the results.
MAE (\autoref{equ:error_indicators_mae}) is less sensitive to this, though not immune.

\begin{figure}
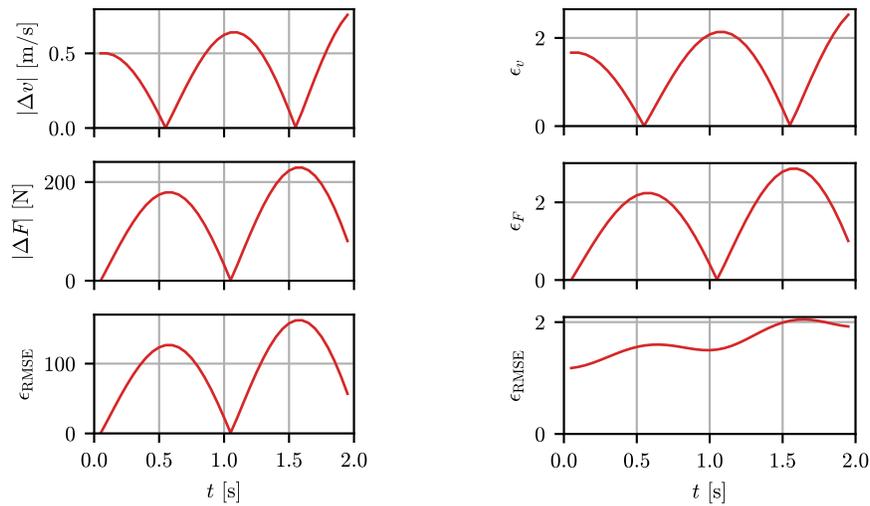

    \centering
    \begin{subfigure}{0.45\textwidth}
        \centering
        \includesvg[width=1\textwidth]{figures/error_indicator_unnormalized.svg}
        \caption{Unnormalised individual errors.}
        \label{fig:error_indicator_unnormalized}
    \end{subfigure}
    \hfill
    \begin{subfigure}{0.45\textwidth}
        \centering
        \includesvg[width=1\textwidth]{figures/error_indicator_normalized.svg}
        \caption{Normalised individual errors.}
        \label{fig:error_indicator_normalized}
    \end{subfigure}
    \caption{%
        Local errors in the outputs for the undamped harmonic oscillator co-simulation example from \autoref{subsec:errorestimation_example}.
        (\subref{fig:error_indicator_unnormalized})~Notice how the na\"ive use of the root-mean-squared error (RMSE) for aggregation into a scalar error indicator almost completely disregards error contributions from the velocity output $|\Delta v|$ in favour of the numerically much larger contributions from the force output $|\Delta F|$.
        (\subref{fig:error_indicator_normalized})~This can be remedied with an appropriate normalisation of the individual error contributions, but this may not always be straightforward in practice.%
    }
    \label{fig:error_indicator}
\end{figure}

We have already encountered an example of this in the simple toy system from \autoref{subsec:errorestimation_example}:
Compare the numeric scale of the force in Newton to that of the resulting velocity expressed in meters per second.
Even worryingly large errors in the velocity (or displacement) will be drowned out by tiny errors in the force, unless the numeric values are normalised properly; see \autoref{fig:error_indicator}.
Such normalisation can be achieved by using adequate absolute and relative tolerances, as done in \autoref{equ:error_indicator_normalize}, or by using aggregation functions that are more robust to large numeric contributions, such as quantiles.
Some error estimators sidestep the issue entirely;
see the work of \textcite{inci2023error} on serial co-simulation for an example.

\subsubsection{Non-locality of errors}
\label{subsec:error_recommendations_nonlocality}

This also highlights another issue with co-simulation:
Local error estimators can help identify local issues, but because of complex interactions between subsystems, the problem can lie somewhere else entirely.
Because of that very complexity and the infinitely many diverse ways in which someone can set up a co-simulation from individual subsystems it is very difficult to provide general advice.
But, at the very least, it is worth keeping this point in mind when analysing coupling errors and when trying to make sense of unexpected behaviour.


\section{Step size control}
\label{sec:stepsizecontrol}

In the previous section, we learned a few ways to construct an error indicator $\epsilon$ for a co-simulation.
Furthermore, we learned that $\epsilon$ generally depends on the macro time step size $\Delta t$, since increasing the step size will tend to increase coupling errors.
Because simulation speed---the RTF, that is---also increases with $\Delta t$, there is a positive correlation between errors and simulation speed.
The more accurate we want our results to be, the slower our simulations will run.

Quite often, what we want is for our simulations to run as fast as possible, while keeping errors below a certain threshold.
In other words, we want to keep $\epsilon$ as close as possible to, but ideally not above, a certain threshold value.
It is conventional to take this value to be unity, so $\epsilon \lesssim 1$, and state our accuracy requirements in terms of the absolute and relative tolerances.
One way to achieve this is to evaluate the error indicator for each macro time step, and if $\epsilon > 1$, roll back to the previous synchronisation point and retry with a shorter step size.
Repeat if the error indicator is still too large.

Here, however, we are interested in the situation where we have minimal model interfaces, so rollback is off the table.
All we can do then is to control the size of the \emph{next} time step, and we may have to accept that the error indicator sometimes exceeds the threshold.
In other words, we can only aim for $\epsilon \approx 1$.

\begin{figure}
    \centering
    \includesvg{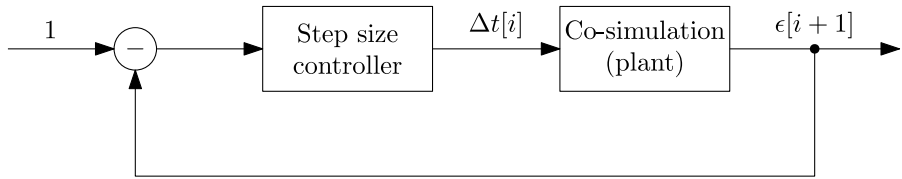}
    \caption{Block diagram of a step size control loop.}
    \label{fig:step_size_control_block_diagram}
\end{figure}

This can be viewed as a discrete control problem.
In control-theoretical terms, a co-simulation is then a \emph{single-input single-output}, or SISO, system.
The input, or \emph{control signal}, is the macro time step size, while the output that we wish to control is the error indicator.
\autoref{fig:step_size_control_block_diagram} shows this schematically.
In general, we must assume that the plant---the co-simulation, that is---has a nonlinear response to the input.

Designing an optimal controller and proving its stability for a given system requires a certain level of knowledge about the system.
In this article, we are concerned with systems where we have very little such knowledge.
Therefore, we will discuss a simple and robust controller that can be expected to work well in most cases, and which makes minimal assumptions about the system being controlled: a plain proportional--integral (PI) controller.
But before we get to \emph{automatic} step size control, let us review the most basic co-simulation algorithm: the fixed-step algorithm.

\subsection{Basic co-simulation algorithm}

The simplest possible co-simulation algorithm is one where the macro step size is kept fixed throughout the simulation, and where all subsystems share the same synchronisation points.
This is shown in pseudocode in Algorithm \ref{alg:fixed_step}.
It starts by setting the initial values of the simulation time $t$ and the vector $\mathbf{u}$ which contains all subsystem input variable values.
The following loop is thereafter repeated until the desired end time is reached:
First, the input values are transferred to the respective subsystems.
Then, the subsystems are instructed to advance forward in time to the next synchronisation point at $t_\mathrm{next}$.
(Note that to avoid overshooting $t_\mathrm{stop}$, the last step may be shorter than $\Delta t$.)
Upon reaching the next synchronisation point, the output values are retrieved from all subsystems and stored in the vector $\mathbf{y}$.
These are then transferred over to the associated inputs via the connection graph matrix $\mathrm{L}$, which we first saw back in \autoref{subsec:errorestimation_estimation_nepce}, and the loop repeats.
\begin{algorithm}
    \caption{Basic fixed-step co-simulation algorithm.}
    \label{alg:fixed_step}
    \begin{algorithmic}[1]
        \State $t \gets t_\mathrm{start}$
        \State $\mathbf{u} \gets \mathbf{u}_\mathrm{start}$
        \While{$t < t_\mathrm{stop}$}
            \State \Call{SetInputVariables}{$\mathbf{u}$}
            \State $t_\mathrm{next} \gets \min(t + \Delta t, t_\mathrm{stop})$
            \State \Call{DoStep}{$t_\mathrm{next} - t$}
            \State $t \gets t_\mathrm{next}$
            \State $\mathbf{y} \gets$ \Call{GetOutputVariables}{$\,$}
            \State $\mathbf{u} \gets \mathrm{L} \mathbf{y}$
        \EndWhile
    \end{algorithmic}
\end{algorithm}

The advantage of this algorithm is that it is simple to understand, implement, and use.
It has only one parameter, $\Delta t$, which can be tuned to obtain the desired balance between accuracy and speed.
The downside is that this balance can be hard to strike, especially when dealing with black-box, nonlinear subsystems.
But even though the algorithm itself does not make use of an error indicator, the error indicator can still be calculated and used as a check of simulation accuracy and to manually tune the step size.
If $\epsilon \ll 1$ throughout a simulation run, then it may be a sign that the step size can be increased in future runs, and if $\epsilon > 1$ in parts of the run, then the step size should be reduced and the simulation run repeated.
See \autoref{subsec:error_recommendations_large_errors} and \autoref{fig:ecco_stability} for a more in-depth look at how error indicators can help with this task.

\subsection{Automatic step size control}

Algorithm \ref{alg:auto_step} is a modification of Algorithm \ref{alg:fixed_step} which allows for a step size controller to be plugged in.
Specifically, lines \ref{algl:new_delta_t_init}, \ref{algl:new_error_indicator}, and \ref{algl:new_step_size} have been added.
The step size is now dynamic, but we still need to provide a start value.
After performing a macro step and retrieving the output values, the error indicator $\epsilon$ is calculated.
Here, the \emph{ComputeErrorIndicator} placeholder function could stand in for any of the methods we explored in \autoref{subsec:errorestimation_estimation_minimal}.
Thereafter, the size of the next macro time step is calculated on the basis of $\epsilon$.
\begin{algorithm}
    \caption{Co-simulation algorithm with automatic step size control.}
    \label{alg:auto_step}
    \begin{algorithmic}[1]
        \State $t \gets t_\mathrm{start}$
        \State $\Delta t \gets \Delta t_\mathrm{start}$                                     \label{algl:new_delta_t_init}
        \State $\mathbf{u} \gets \mathbf{u}_\mathrm{start}$
        \While{$t < t_\mathrm{stop}$}
            \State \Call{SetInputVariables}{$\mathbf{u}$}
            \State $t_\mathrm{next} \gets \min(t + \Delta t, t_\mathrm{stop})$
            \State \Call{DoStep}{$t_\mathrm{next} - t$}
            \State $t \gets t_\mathrm{next}$
            \State $\mathbf{y} \gets$ \Call{GetOutputVariables}{$\,$}
            \State $\epsilon \gets$ \Call{ComputeErrorIndicator}{$\mathbf{u},\mathbf{y}$}   \label{algl:new_error_indicator}
            \State $\Delta t \gets $ \Call{ComputeNextStepSize}{$\Delta t, \epsilon$}       \label{algl:new_step_size}
            \State $\mathbf{u} \gets \mathrm{L} \mathbf{y}$
        \EndWhile
    \end{algorithmic}
\end{algorithm}

We will now derive a simple step size controller.
Our derivation will closely follow that of \textcite{gustafsson1988pi}, who discuss step size control in the context of numerical solution of ODEs.
Here, we have adapted it slightly to the context of co-simulation.

Let us assume that the error indicator is of polynomial order $p$ in the macro time step size, that is, $\epsilon = \mathcal O(\Delta t^p)$.\footnote{
    For NEPCE, $p=m+1$, where $m$ is the input extrapolation order.
    For the explicit predictor/corrector, $p=m+1$ in the presence of direct feed-through, otherwise $p=m+2$.
    For ECCO, $p=m+2$.
    \cite{sadjina2020energy}
}
At synchronisation point $i$, we then have
\begin{equation}
    \epsilon[i] = \phi[i] \Delta t[i-1]^p, \label{equ:epsilon_i_of_time}
\end{equation}
where $\phi[i]$ depends on the system dynamics and is $\mathcal O(1)$ in $\Delta t$.
Here, we have used the convention that $\Delta t[i]$ is the size of the macro time step which starts at synchronisation point $i$, that is, $\Delta t[i] = t[i+1] - t[i]$.

We now wish to calculate the next step size, $\Delta t[i]$, such that the error indicator is approximately equal to 1 at the next synchronisation point.
If we achieve this, we can assume that the synchronisation rate will have been much faster than changes within the system, so that $\phi[i+1] \approx \phi[i]$.
Thus,
\begin{equation}
    \epsilon[i+1] \approx \phi[i] \Delta t[i]^p. \label{equ:slow_dynamics_assumption}
\end{equation}
Since we are aiming for $\epsilon[i+1] \approx 1$, a good candidate for $\Delta t[i]$ can be found by solving the equation
\begin{equation}
    \phi[i] \Delta t[i]^p = 1.
\end{equation}
Substituting for $\phi[i]$ using \autoref{equ:epsilon_i_of_time}, we get
\begin{equation}
    \Delta t[i] = \epsilon[i]^{-1/p} \Delta t[i-1].
\end{equation}
For increased robustness, it is useful to place lower and upper limits on the value of $\Delta t[i]$, as well as on its rate of change.
With this modification, we get
\begin{subequations}
    \label{equ:i_controller}
    \begin{align}
        \Delta t'[i] &= \epsilon[i]^{-1/p} \Delta t[i-1], \label{equ:dt_prime} \\
        \Delta t[i] &= \min(
            \max(\Delta t'[i], \Delta t_\mathrm{min}, \theta_\mathrm{min} \Delta t[i-1]),
                \Delta t_\mathrm{max}, \theta_\mathrm{max} \Delta t[i-1]),
    \end{align}
\end{subequations}
where $\Delta t_\mathrm{min}$ and $\Delta t_\mathrm{max}$ are the smallest and largest allowed step sizes, respectively, while $\theta_\mathrm{min}$ and $\theta_\mathrm{max}$ are the corresponding limits on the rate of change.

From a control-theoretical perspective, \autoref{equ:i_controller} actually describes an \emph{integrating controller}.
This can be shown more explicitly by manipulating the equations a bit.
First, take the logarithm on both sides of \autoref{equ:dt_prime}:
\begin{equation}
    \log \Delta t'[i] = -\frac{1}{p} \log \epsilon[i] + \log \Delta t[i-1].
\end{equation}
Now, let $I[i] = p \log \Delta t[i]$ and $I'[i] = p \log \Delta t'[i]$, and define $e[i] = -\log\epsilon[i]$ and $k_I = 1/p$.
Then, we find that
\begin{subequations}
    \label{equ:i_controller_spoonfed}
    \begin{align}
        I'[i] &= I[i-1] + e[i], \label{equ:i_controller_spoonfed_integration} \\
        \log \Delta t'[i] &= k_I I'[i], \label{equ:i_controller_spoonfed_control} \\
        \Delta t[i] &= \min(
            \max(\Delta t'[i], \Delta t_\mathrm{min}, \theta_\mathrm{min} \Delta t[i-1]),
                \Delta t_\mathrm{max}, \theta_\mathrm{max} \Delta t[i-1]), \label{equ:i_controller_spoonfed_limitation} \\
        I[i] &= I'[i] + k_I^{-1} (\log \Delta t[i] - \log \Delta t'[i]). \label{equ:i_controller_spoonfed_anti_windup}
    \end{align}
\end{subequations}
If we interpret $e$ as a control error to be minimised, then these equations clearly describe an integrating controller where $\log \Delta t$ is the control variable and $k_I$ is the integrator gain:
\autoref{equ:i_controller_spoonfed_integration} is the integration step, \autoref{equ:i_controller_spoonfed_control} is the control law, \autoref{equ:i_controller_spoonfed_limitation} is a limitation of the control variable, and \autoref{equ:i_controller_spoonfed_anti_windup} constitutes a form of anti-windup.
(Note that the last equation follows quite trivially from the previously-stated relations between $I$ and $\Delta t$ and between $I'$ and $\Delta t'$.)

\textcite{gustafsson1988pi} point out that this controller performs poorly in many circumstances.
In particular, it is prone to oscillations.
They make two observations concerning the origin of such oscillations: firstly, that the integral gain $1/p$ is too large; secondly, that a pure integrating controller in itself has poor stabilising properties.
Therefore, they suggest replacing it with a proportional--integral (PI) controller, and letting both the proportional and integral gains be tunable parameters.

Introducing $k_P$ as the proportional gain, the PI version of \autoref{equ:i_controller_spoonfed} is
\begin{subequations}
    \label{equ:pi_controller_spoonfed}
    \begin{align}
        \mathcal I'[i] &= \mathcal I[i-1] + k_I e[i], \label{equ:pi_controller_spoonfed_integration} \\
        \log \Delta t'[i] &= k_P e + \mathcal I'[i], \label{equ:pi_controller_spoonfed_control} \\
        \Delta t[i] &= \min(
            \max(\Delta t'[i], \Delta t_\mathrm{min}, \theta_\mathrm{min} \Delta t[i-1]),
                \Delta t_\mathrm{max}, \theta_\mathrm{max} \Delta t[i-1]), \label{equ:pi_controller_spoonfed_limitation} \\
        \mathcal I[i] &= \mathcal I'[i] + \log \Delta t[i] - \log \Delta t'[i]. \label{equ:pi_controller_spoonfed_anti_windup}
    \end{align}
\end{subequations}
Note that we have now absorbed the $k_I$ coefficient into the error sum $\mathcal I$ to maintain a simple form for the anti-windup equation, which is now less trivial (and somewhat unconventional for a PI controller) because it also incorporates the effects of the proportional term.

By reversing the manipulations we performed to get from \autoref{equ:i_controller} to \autoref{equ:i_controller_spoonfed}, we can also write \autoref{equ:pi_controller_spoonfed} in a more compact form:
\begin{subequations}
    \label{equ:pi_controller_compact}
    \begin{align}
        \Delta t'[i] &= \epsilon[i]^{-k_P-k_I} \epsilon[i-1]^{k_P} \Delta t[i-1], \label{equ:pi_controller_compact_unlimited} \\
        \Delta t[i] &= \min(
            \max(\Delta t[i]', \Delta t_\mathrm{min}, \theta_\mathrm{min} \Delta t[i-1]),
                \Delta t_\mathrm{max}, \theta_\mathrm{max} \Delta t[i-1]).
    \end{align}
\end{subequations}

\subsection{Implementing and configuring the step size controller}
\label{sec:step_size_control_implementation}

While the PI controller is often presented in the form shown in \autoref{equ:pi_controller_compact} \cite{gustafsson1988pi,gustafsson1991control,soderlind2002automatic}, we recommend something more like \autoref{equ:pi_controller_spoonfed} for practical numerical implementations, mainly because it is easier to read and understand.
And, perhaps surprisingly, it is no less computationally efficient than the compact form.
The corresponding pseudocode is shown in Algorithm \ref{alg:pi_controller}.
\begin{algorithm}
    \caption{
        Pseudocode for a PI step size controller based on \autoref{equ:pi_controller_spoonfed}.
        The error sum $\mathcal I$ must persist between function calls; all other variables are local.
        The \emph{ComputeNextStepSize} function can be called from a co-simulation master algorithm as shown in Algorithm \ref{alg:auto_step}.
    }
    \label{alg:pi_controller}
    \begin{algorithmic}[1]
        \State $\mathcal I \gets 0$
        \Function{ComputeNextStepSize}{$\Delta t_\mathrm{old}, \epsilon$}
            \State $e \gets -\log \epsilon$
            \State $\mathcal I' \gets \mathcal I + k_I e$
            \State $\ell \gets k_P e + \mathcal I'$
            \State $\Delta t' \gets \exp(\ell)$
            \State $\Delta t \gets \min(
                \max(\Delta t', \Delta t_\mathrm{min}, \theta_\mathrm{min} \Delta t_\mathrm{old}),
                \Delta t_\mathrm{max}, \theta_\mathrm{max} \Delta t_\mathrm{old})$
            \State $\mathcal I \gets \mathcal I' + \log \Delta t - \ell$
            \State \Return{$\Delta t$}
        \EndFunction
    \end{algorithmic}
\end{algorithm}

The control algorithm has a few parameters that need to be specified.
First and foremost, the controller gains $k_P$ and $k_I$ must be chosen appropriately.
The optimal values for these parameters will depend on the overall system dynamics and on individual subsystem characteristics such as input extrapolation order and integration order.
There is much to be said on the topic of PI controller tuning, but we will not delve into control theory here, and instead refer readers to standard textbooks on the matter.
However, we will suggest the following as a starting point for such tuning:
\begin{equation}
    k_P = \frac{0.4}{p}, \qquad k_I = \frac{0.3}{p}.
\end{equation}
These were first suggested in the context of step-size control for numerical solution of ODEs \cite{gustafsson1991control}, but have been shown to work well for co-simulation benchmark cases too \cite{sadjina2016energy,sadjina2020energy}.

Secondly, since we don't have an error estimate before we have performed the first step, $\Delta t[0]$ must be chosen manually.
It should be set to a small value to ensure that the assumption in \autoref{equ:slow_dynamics_assumption} isn't violated from the outset.
A conservative choice is
\begin{equation}
    \Delta t[0] = \Delta t_\mathrm{min}.
\end{equation}
If the initial errors are small, the controller will immediately start to increase the step size towards a more appropriate value.

There are four different parameters that limit the step size and step size changes in \autoref{equ:pi_controller_spoonfed_limitation}.
Of these, $\Delta t_\mathrm{max}$ is arguably the most important.
Using an appropriate value for it can prevent the controller from venturing into territory where the error indicator becomes unreliable, as discussed in \autoref{subsec:error_recommendations_large_errors}.
$\Delta t_\mathrm{min}$, on the other hand, could in principle be set to zero---at least if simulation accuracy was all that mattered.
However, in most cases we have a limit to how long we wish to wait for a simulation to complete, so there is a tipping point where speed trumps accuracy.
$\Delta t_\mathrm{min}$ allows us to specify this tipping point.
This is especially useful during controller tuning, where the wrong parameters can easily cause the algorithm to settle on the minimum step size.

The rate limiting parameters $\theta_\mathrm{min}$ and $\theta_\mathrm{max}$ can increase the robustness of the controller by preventing it from over-eagerly making large step size changes, which could happen if the simulated system is highly nonlinear or if the error indicator contains noise.
Some sources suggest values in the ranges $\theta_\mathrm{min} \in [0.2, 0.5]$ and $\theta_\mathrm{max} \in [1.5, 5]$, and these ranges may well be a reasonable starting point \cite{schierz2012cosimulation,hairer1993solving}.
It should be noted that the sources in question discuss \emph{re-stepping} algorithms for co-simulation and ODE solving, respectively.
In such algorithms, if the error indicator for a step is too large, the step is rolled back and retaken with a shorter step size.
This is computationally very costly, and such algorithms tend to be conservative in order to reduce the number of rejected steps.
It is arguably a good idea to be conservative in non-restepping algorithms too, precisely since they are unable to correct for previous errors, but this is an area where further investigation is warranted.

For the same reason, it is quite common for re-stepping algorithms to multiply the candidate step size by a ``safety factor'' $\alpha \in [0.8, 0.9]$.
With this, \autoref{equ:pi_controller_compact_unlimited} would be modified to
\begin{equation}
    \Delta t'[i] = \alpha \epsilon[i]^{-k_P-k_I} \epsilon[i-1]^{k_P} \Delta t[i-1].
\end{equation}
However, this simply introduces a bias in the controller which makes it aim for some value of $\epsilon$ which is slightly lower than 1.
In the non-restepping algorithm, where we do accept that the value of $\epsilon$ sometimes exceeds 1, this seems rather pointless.\footnote{
    We ourselves are guilty of unthinkingly including the safety factor in \textcite{sadjina2016energy}.
}
Instead, we recommend lowering the tolerances if greater accuracy is needed.

\subsection{An example}

\autoref{fig:quarter_car_step_size_control} demonstrates the
effect of implementing automatic step size control according to
Algorithm~\ref{alg:auto_step} and Algorithm~\ref{alg:pi_controller} for the
\emph{quarter car} system.
The quarter car is a simplified chassis--suspension--wheel system modelled as two
coupled harmonic oscillators and is often used as a simulation benchmark case.
Here, we separated it into two subsystems as shown in
\autoref{fig:quarter_car}, connected via force--velocity coupling.
See \textcite{schierz2012cosimulation} and \textcite{sadjina2016energy} for
in-depth descriptions of the setup.
We used the same model parameters as both those papers, namely
$m_c = \SI{400}{\kilogram}$, $m_w = \SI{40}{\kilogram}$,
$k_c = \SI{1.5e4}{\newton\per\meter}$, $k_w = \SI{1.5e5}{\newton\per\meter}$,
and $d_c = \SI{1e3}{\newton\second\per\meter}$.
\begin{figure}
    \centering
    \includesvg[width=\textwidth]{figures/quarter_car_step_size_control.svg}
    \caption{%
        Quarter car benchmark results demonstrating the implementation of the fixed-step size Algorithm~\ref{alg:fixed_step} (left) and the automatic step size Algorithm~\ref{alg:auto_step} (right).
        An error indicator can be used to track the accuracy of the simulation as it progresses (top).
        This allows for automatic step size control (second row) to help keep the
        coupling errors within acceptable bounds (bottom two rows).%
    }
    \label{fig:quarter_car_step_size_control}
\end{figure}
\begin{figure}
    \centering
    \includesvg{figures/quarter_car.svg}
    \caption{%
        A schematic of the quarter car benchmark system.
        The bar on the left indicates how the system was split into subsystems
        for the purpose of co-simulation, with $S_1$ representing the chassis
        and $S_2$ containing the suspension and wheel.
    }
    \label{fig:quarter_car}
\end{figure}

The PI step size controller was configured using the controller gains
$k_p = 0.4$ and $k_I = 0.3$ as suggested above;
the step size bounds $\Delta t_\mathrm{min} = \SI{1e-4}{\second}$ and
$\Delta t_\mathrm{max} = \SI{1e-2}{\second}$;
and the rate limiting parameters $\theta_\mathrm{min} = 0.2$ and
$\theta_\mathrm{max} = 1.5$.
An RSME error indicator (\autoref{equ:error_indicators_rmse}) was used with the
NEPCE local input error estimation calculated from~\autoref{equ:error_nepce_zoh}.
The input variables are $u_1$, the force from the wheel--suspension system
($S_2$) on the chassis ($S_1$), and $u_2$, the velocity of the chassis as input
to $S_2$.
The fixed-step-size comparison simulation, based on
Algorithm~\ref{alg:fixed_step}, was performed using a step size of
$\Delta t = \SI{1e-3}{\second}$.

We mentioned in \autoref{sec:error_indicator} that it can be challenging to
choose appropriate tolerance values.
This is true even for a simple system such as this.
Here, we have four tolerances that we need to define:
a relative and an absolute tolerance for both the force error and the velocity
error.
In addition, we have a system that is characterised by initially large
oscillations which are subsequently damped out.
As both force and velocity approach zero at large $t$, the absolute tolerances
will eventually dominate the relative tolerances, and this might happen at
different times for the two quantities.
Different tolerance values will weight accuracy in the dynamic part and the
steady-state part differently, and they will weight accuracy in the force and
velocity differently.

We wanted to obtain approximately equal relative accuracy in both quantities,
and we wanted to emphasise accuracy in the highly dynamical early phase. Here is
the procedure we followed to set the tolerances:
\begin{enumerate}
    \item Look at the quantities whose errors are being estimated to determine
        their typical magnitudes.
        What is ``typical'' is to some extent a judgment call, since the
        quantities vary a lot over the period of interest.
        By considering \autoref{fig:quarter_car_inputs}, we deemed that
        $\bar u_1 = \SI{1e3}{\newton}$ and
        $\bar u_2 = \SI{0.3}{\meter\per\second}$ were sensible scales.
    \item Use a single relative tolerance $\sigma$ for both quantities, so
        $\sigma_{u_1} = \sigma_{u_2} = \sigma$, and construct the absolute
        tolerances as $\delta_{u_k} = \sigma_{u_k} \bar u_k$.
        Then, the normalised error indicator from
        \autoref{equ:error_indicator_normalize} takes the form
        \begin{equation}
            \epsilon_{u_k}[i] = \frac{\Delta u_k[i]}{\sigma (\bar u_k + u_k[i])}.
        \end{equation}
        Now, we have reduced the number of ``tolerance parameters'' to three,
        where two are physically motivated and one refers to the desired
        accuracy.
    \item Adjust the relative tolerance until satisfied with the results.
\end{enumerate}
Here, we aimed to show that an automatic step size controller can yield greater
accuracy at the same computational cost compared to the fixed-step algorithm.
Therefore, we adjusted $\sigma$ so that the total number of time steps was
approximately the same in the two cases (i.e., around 4000).
Thus, we ended up with $\sigma = 2\times 10^{-3}$.
\begin{figure}
    \centering
    \includesvg[width=\textwidth]{figures/quarter_car_inputs.svg}
    \caption{%
        The values of the two input variables in the quarter car system, as
        simulated using the fixed-step algorithm, plotted on a logarithmic
        scale.
    }
    \label{fig:quarter_car_inputs}
\end{figure}

For more examples, we refer the reader to \textcite{sadjina2016energy}, where
the same case as well as a nonlinear variant is studied in more detail, and
to \textcite{sadjina2024energy}, where we deal with the slightly more complex
\emph{quarter truck} system.

\subsection{Research needs}

These last few subsections may have left readers with a background in control theory somewhat unsatisfied.
We have only gestured vaguely at some control-theoretic concepts and borrowed a few equations from ODE solver theory.
While this was deliberate given the topic and scope of this paper, it is worth mentioning that the academic literature at large is also rather thin in this area.

A couple of contributions are worth mentioning.
\textcite{arnold2013error} perform a convergence analysis which shows that global coupling errors over a finite time interval are bounded in terms of local errors, provided certain conditions are fulfilled.
Thus, they give a theoretical justification for the use of local coupling errors as the controlled output in a step size control loop like the one shown in \autoref{fig:step_size_control_block_diagram}.
In a more recent paper, \textcite{glumac2023defect} present a proof that coupling errors in a co-simulation can be controlled via the macro step size when the subsystems are coupled ODE systems.
\textcite{skjong2019numerical} derive a global stability criterion for co-simulated systems more generally, and discuss conditions under which stability can be guaranteed.
However, they do not explicitly discuss the implications this may have for step size control.

We believe there is a need for further research which treats a co-simulation as a SISO system for the purpose of studying its stability characteristics and devising optimal control systems for it.
Given the black-box, nonlinear nature of many co-simulated systems, this may be tricky.
Perhaps ideas could be taken from sampled systems theory and system identification to suggest new ways to characterise co-simulations and subsystems, and to link them to conditions under which specific controller types are applicable.
It would also be interesting to see some of the more recent research on ODE solver step size control carried over to co-simulation, for example methods based on reinforcement learning \cite{dellnitz2021efficient} or deep learning \cite{liu2020hierarchical}.


\section{Conclusion and recommendations}
\label{sec:discussion}

In this paper, we have attempted to show that sophisticated error estimation and robust step size control is possible even in co-simulations where we have minimal information about the subsystems, and where rollback of macro time steps is not an option.
The methods we have presented are not new, nor are they all our own work, but we believe it is the first time they have been reviewed and presented together within this framing, along with straightforward implementation guidelines.
Our hope is that the article will help both researchers and engineers to get ``up and running'' with high-performance, accurate, \emph{collaborative} co-simulations with minimal friction.

To close, we offer some advice based on our own experience in developing subsystem models and collaborating on co-simulations, gained over a decade of industrial research and innovation projects.
First, concerning error estimation and the more general problem of detecting when a simulation has gone awry and figuring out what to do about it:
\begin{description}
    \item[Know your system.]
        This may seem obvious, but it's worth pointing out:
        You should have a clear expectation of the qualitative outcome of your simulation, even if the quantitative details are hard to predict (which is probably why you're running a simulation in the first place).
    \item[Know your subsystems.]
        Even when you only have access to subsystem inputs and outputs, you need to know what the valid inputs are, that is, each subsystem's \emph{domain of validity}, and what range of outputs to expect.
        (Ideally, this should be part of the subsystem's documentation.)
    \item[Know your connections.]
        In particular, know which connections represent physical interactions, and which represent digital information channels.
        Only the former should be used in the calculation of error estimates.
    \item[Estimate errors.]
        Local error estimates are an invaluable tool for debugging simulations, even when they're not used for automatic step size control.
        As \autoref{subsec:errorestimation_estimation_minimal} shows, there are several methods to choose from even in the minimal-interface case.
        Try them out and pick the one(s) which work best for your particular system.
\end{description}
Next, on the topic of step size control:
\begin{description}
    \item[Keep it simple.]
        It is often better to use a robust, conservative controller than a complex one tailored for a specific type of problem.
        The reason is that, often, more time is spent setting up a simulation than actually running it.
        Exceptions to this rule might be if your work involves very long-running simulations, systems which are highly prone to instability, or very specific and strict requirements with regards to speed or accuracy.
    \item[Know your controller.]
        You should have a basic understanding of how your step size controller works, so that you know which factors contribute to driving the step size up or down.
    \item[Establish a baseline.]
        As we have seen, the magnitude of coupling errors depend on the step size, and errors can lead to a distortion of the physics inside the subsystems.
        Thus, a step size controller in some sense becomes a dynamical part of the system, and its effects can be difficult to disentangle from the actual physics of the simulation.
        To establish a baseline, switch to a simple fixed-step algorithm with a small step size and compare the results.
    \item[Start conservatively.]
        Whether using manual or automatic step size control, begin with a small step size.
        In the manual case, increase it while keeping track of errors, and when $\epsilon \gg 1$, dial back a bit.
        A good automatic step size controller should do this for you, but it needs to be given a sufficiently small initial step size so that the assumptions that underlie it hold, as we discussed in \autoref{sec:step_size_control_implementation}.
    \item[Monitor step sizes and error estimates.]
        If the step size algorithm has settled on the minimum step size, but the error estimates keep increasing, then the system is most likely numerically or dynamically unstable (or your $\Delta t_\mathrm{min}$ is too large).
        Oscillations in the error estimates or the step sizes can stem from the tuning of the controller (try lowering the gains) or can be observed when the simulation is close to, or exceeds, the critical step size beyond which it becomes unstable (try lowering $\Delta t_\mathrm{max}$ or decrease tolerances).
\end{description}

\subsection*{Code and data}

The Python code and data used to produce the plots in this article are available
at \url{https://doi.org/10.60609/n0a1-2t41}.

\subsection*{CRediT author statement}

\textbf{Lars T. Kyllingstad:} Conceptualization, Funding acquisition, Investigation, Writing -- original draft.
\textbf{Severin Sadjina:} Conceptualization, Funding acquisition, Investigation, Visualization, Writing -- original draft.
\textbf{Stian Skjong:} Conceptualization, Funding acquisition, Investigation, Writing -- review \& editing, Project administration.

\subsection*{Acknowledgements}
This work was supported by the project \emph{SEACo:\ Safer, Easier, and more Accurate Co-simulations}, funded by the Research Council of Norway under grant number 326710.


{
    \raggedright
    \printbibliography
}

\end{document}